 \documentclass[preprint, review,3p,times, 10pt]{elsarticle}

\usepackage{amssymb}
\usepackage{pgf-pie}
\usepackage{comment}
\usepackage{multirow}
\usepackage{makecell}
\usepackage[hyphens]{url}
\usepackage{hyperref}
\usepackage{pdfpages}
\hypersetup{colorlinks=true,breaklinks=true}

\journal{'}

\begin{document}

\begin{frontmatter}


\title{From Lab to Pocket: A Novel Continual Learning-based Mobile Application for Screening COVID-19}

\author[1]{Danny Falero}\ead{dfalero@csu.edu.au}

\author[1]{Muhammad Ashad Kabir\corref{mycorrespondingauthor}}
\address[1]{School of Computing, Mathematics and Engineering, Charles Sturt University, Bathurst, NSW 2795, Australia}
\cortext[mycorrespondingauthor]{Corresponding author}
\ead{akabir@csu.edu.au}

\author[2]{Nusrat Homaira}\ead{n.homaira@unsw.edu.au}
\address[2]{School of Clinical Medicine, UNSW, Sydney, NSW 2031, Australia}
\begin{abstract}
Artificial intelligence (AI) has emerged as a promising tool for predicting COVID-19 from medical images. In this paper, we propose a novel continual learning-based approach and present the design and implementation of a mobile application for screening COVID-19. Our approach demonstrates the ability to adapt to evolving datasets, including data collected from different locations or hospitals, varying virus strains, and diverse clinical presentations, without retraining from scratch. We have evaluated state-of-the-art continual learning methods for detecting COVID-19 from chest X-rays and selected the best-performing model for our mobile app. We evaluated various deep learning architectures to select the best performing one as a foundation model for continual learning. Both regularization and memory-based methods for continual learning were tested, using different memory sizes to develop the optimal continual learning model for our app.
DenseNet161 emerged as the best foundation model with 96.87\% accuracy, and Learning without Forgetting (LwF) was the top continual learning method with an overall performance of 71.99\%. The mobile app design considers both patient and doctor perspectives. It incorporates the continual learning DenseNet161 LwF model on a cloud server, enabling the model to learn from new instances of chest X-rays and their classifications as they are submitted. The app is designed, implemented, and evaluated to ensure it provides an efficient tool for COVID-19 screening. The app is available to download from \url{https://github.com/DannyFGitHub/COVID-19PneumoCheckApp}.
\end{abstract}




\end{frontmatter}


\section{Introduction}\label{lbl:introduction}
The application of artificial intelligence (AI) in medical imaging has advanced rapidly in recent years~\citep{ting2018ai,giger2018machine,pesapane2018artificial}
Machine learning (ML), particularly deep learning (DL) based processing of chest X-ray images (CXRs) has yielded remarkable success in the automated classification of lung pathologies, including COVID-19~\citep{Wang2019,Bullock2020,qin2021tuberculosis}. DL models have achieved exceptional performance, surpassing the capabilities of board-certified radiologists in specific, well-defined tasks~\citep{Rajpurkar2017}.
Automated DL systems for the screening of pneumonia have the potential to enable early intervention and provide support for clinical decision-making~\citep{wang2021deep}. Although reverse transcription polymerase chain reaction (RT–PCR) is the preferred method for diagnosing COVID-19, imaging can serve as a complementary tool to improve diagnostic accuracy. Additionally, chest X-ray (CXR) abnormalities can sometimes be detected in patients who initially test negative with RT–PCR~\citep{wong2020frequency}.

Machine learning models used for detecting COVID-19 in chest X-rays may struggle to generalise across different virus strains or datasets~\citep{cao2024reinvestigating,paul2020generalizability,Panday2020}. This challenge occurs when models trained on a specific dataset—such as one containing images from patients with a particular strain or from a specific hospital—encounter data from a different strain or a new source~\citep{roberts2021common}. This can lead to a drop in the model's accuracy because it may have learned to identify confounding factors, such as image artifacts, rather than focusing on medically relevant features of the disease. Therefore, a model successful with one dataset or source might not work effectively for another due to these variations~\citep{sahiner2023data}. Addressing this challenge using conventional machine learning requires gathering and retraining on new data, which can be cumbersome and impractical. Furthermore, conventional machine learning models require the entire dataset for retraining when faced with new strains or clinical pictures, incurring substantial costs. Therefore, there is a need for adaptive and efficient diagnostic tools that can handle evolving data patterns and new variants of the virus.

Continual learning, an innovative machine learning approach, offers significant advantages over traditional task sequential machine learning~\citep{Hadsell2020}. It allows models to adapt continuously to changes in the clinical picture, symptoms, and infection rates associated with new COVID-19 strains. Continual learning networks can incrementally learn from new COVID-19 patient diagnosis confirmations, providing up-to-date, highly accurate, and instant chest X-ray diagnoses, ultimately alleviating the strain on public health systems. Furthermore, mobile applications equipped with machine learning capabilities have the potential to revolutionize COVID-19 diagnosis and patient monitoring. They can reduce the demand for clinical testing, and provide rapid and efficient assessments using CXR. As the number of smart devices grows and hardware advances, deploying machine learning models on mobile devices becomes increasingly important to minimize latency and provide real-time analysis.

This study aimed to develop a mobile application for the continual screening of COVID-19 in patients using chest X-ray images. Recognizing the limitations of conventional machine learning, we adopted an incremental continual learning approach. The approach began with creating a foundational model trained using traditional sequential supervised learning on labeled chest X-rays. Subsequently, we transitioned to a continual learning model by transferring knowledge from the foundational model that demonstrated the best performance. 
The continual learning model was then trained using the most effective and established continual learning strategies, settings, and methods, focusing on learning from chest X-rays labeled by medical professionals to provide COVID-19 classifications for chest X-ray images. Following the model's development, it was integrated into a COVID-19 screening mobile application designed for rapid diagnosis. 

Our experimental evaluations demonstrated the effectiveness of our continual machine learning approach, achieving an accuracy of 94.44\% in COVID-19 classifications. This empowers users to efficiently screen for COVID-19 through mobile applications developed using this model. The model's adaptive nature, facilitated by continual learning, ensures its capability to evolve alongside changing clinical presentations. This adaptation enables medical professionals to train a model capable of detecting COVID-19 based on the currently presented conditions.
The key contributions of this paper are as follows:
\begin{itemize}
\item We have evaluated state-of-the-art deep learning models for COVID-19 chest X-ray classification to identify the best model to be used as a foundation model for continual learning.

 \item We have thoroughly examined state-of-the-art continual learning strategies and methods for classifying COVID-19 X-rays in an incremental scenario.

 \item We have proposed an architecture for a mobile app-based COVID-19 screening system. This architecture is designed to cater to the needs of patients, doctors, and researchers, and to incorporate a continual machine learning feature.

 \item We have designed and implemented an Android mobile app for COVID-19 screening and rigorously evaluated its performance.
\end{itemize}

The paper is organized as follows. 
Section~\ref{lbl:background} provides background information laying the foundation for this study.
Section~\ref{lbl:relatedwork} reviews state-of-the-art research relevant to our study. Section~\ref{lbl:methodology} presents the methodology used in our study. Section~\ref{lbl:experiment} presents the experimental evaluation to assess the effectiveness of our continual learning approach. The design, implementation and evaluation of the mobile application are discussed in Section~\ref{lbl:app}. Finally, Section~\ref{lbl:conclusion} concludes the paper and outlines future research directions.

\section{Background}\label{lbl:background}

\subsection{Traditional Machine Learning vs. Continual Learning}

Traditional machine learning models, including support vector machines and decision trees~\citep{Barstugan2020}, as well as deep learning models such as convolutional neural networks~\citep{Oh}, typically follow a training approach involving a single iteration with a fixed dataset of known size. Once trained, these models are used for inference in various applications~\citep{Erickson2021}. However, when retraining is needed, it requires access to both the existing and new data. This retraining process imposes significant demands on data accessibility, availability, storage, and computational resources, as it necessitates rerunning the entire dataset through the training procedure to incorporate the newly added data~\citep{Medera2009}.

Continual learning~\citep{Li2021}, often referred to as lifelong learning~\citep{Hou2018}, incremental learning~\citep{Gepperth2009}, or online learning~\citep{Wong2015}, presents a paradigm distinct from traditional machine learning. While traditional machine learning (i.e., non-continual learning) is typically performed in a single session using a static dataset, continual learning extends training over multiple sessions, allowing adaptation to an indefinite and dynamic dataset~\citep{Hong}. The primary objective of continual learning is to enable models to continuously learn from new data instances without erasing previously acquired knowledge~\citep{Forgetting2018}. A critical challenge in continual learning is catastrophic forgetting, where there is a sharp decline in performance on previously learned tasks after acquiring new knowledge~\citep{French1999,Hasselmo2017}. Continuous learning methodologies aim to prevent such forgetting and address intransigence, which hinders adaptation to new tasks~\citep{Chaudhry2018}. 
To mitigate catastrophic forgetting, three categories of solutions have been identified, as explored by~\citet{Masana2020}: (i) Regularization-based solutions focus on minimizing the impact of new tasks on the weights essential for previous tasks, (ii) Exemplar-based solutions involve storing a limited set of examples and relearning them regularly, effectively preventing forgetting through methods such as replay, and (iii) Task-recency bias minimization solutions aim to reduce the bias towards recently learned instances, thereby preserving the knowledge acquired from earlier tasks.

Continual learning offers three distinctive advantages that traditional machine learning approaches often lack. Firstly, it promotes resource efficiency by eliminating the need for recurrent training from the ground up when new data becomes available~\cite{Polikar2001,Farooq2020}. Continual learning allows a model to seamlessly adapt and train on new data as needed, beyond mere inference~\cite{Lesort2020}. This adaptability is crucial for maintaining classification efficiency, especially in scenarios where new variants of COVID-19 emerge, necessitating rapid learning. Secondly, continual learning reduces the need to store all training data, resulting in lower memory usage by limiting the volume of data that needs to be retained~\cite{Pomponi2020}. This reduction in data storage is crucial for safeguarding the privacy of sensitive COVID-19 patient data. Given the stringent regulations governing data storage in health applications~\cite{McClure2018}, avoiding the storage of patient data on servers ensures compliance and protects against unauthorized access. Lastly, continual learning mirrors human learning processes~\cite{Su2022,Hadsell2020,Flesch2018}. This characteristic is invaluable when training models to perform at a level that matches or surpasses human experts' interpretation of medical data~\cite{Murphy2022}. The synergy between continual learning and human-like learning mechanisms enhances performance and interpretation in the medical field. 

In a nutshell, continual learning delivers resource efficiency, data privacy protection, and human-like learning, making it a promising approach for addressing the dynamic challenges posed by COVID-19 and the broader field of medical data analysis.

\subsection{Transfer Learning vs. Continual Learning}

Transfer learning, a valuable technique in deep learning, involves using sections of a pre-trained model to improve the performance of a new model by leveraging the pre-existing model weights~\cite{Tan2018,Panigrahi2021,Zhuang2021}. This technique is instrumental when building a new model, as parts of a pre-trained model can enhance the new model's performance. The approach often involves retaining the trained feature extraction layers and substituting the classifier with a new one tailored to the classes in the new dataset. Fine-tuning is an instance of transfer learning, where a model trained on a large dataset is adapted to a smaller one by adjusting the last-layer weights~\cite{Rajpura2018}.

While transfer learning is used within the strategies of continual learning~\cite{Thrun1995}, it is crucial to distinguish that mere repetition of transfer learning does not equate to continual learning. Transfer learning involves leveraging parts of a model initially trained for a specific task to address a new task~\cite{Kocer2010}. In contrast, continual learning encompasses training a single model across an indefinite or unknown number of tasks and classes, anticipating an infinite dataset size~\cite{Yang2017}. In essence, continual learning embodies an iterative process where a model is trained and concurrently facilitates inference on an ever-expanding dataset over an extended period, a concept often referred to as ``lifelong learning"~\cite{Parisi2019}.

\subsection{Continual Learning Paradigms}
Continual learning draws upon the foundations of traditional machine learning paradigms, which encompass four key learning approaches: supervised~\cite{Thrun1996,R.R2013,Wiwatcharakoses2021}, semi-supervised~\cite{Chen2020,Luo2022}, unsupervised~\cite{Ott2012,Sharma2012,R.R2013,Liang2017}, and reinforcement learning~\cite{Barreto2012,Ammar,Wu2019,Qian2021}. These paradigms are equally applicable in the context of continual learning, mirroring their roles in traditional machine learning.
Supervised learning, a cornerstone of machine learning~\cite{Li2020}, is used when labeled data are available. The model is tasked with learning from these labeled data to construct a classifier that can classify unlabeled or test data into predefined classes~\cite{Li2020}.
Semi-supervised learning is used when the available labeled data is limited~\cite{Lechat2020}. In this approach, the labeled data is used to predict artificial labels for the unlabeled data, optimizing the learning process.
Unsupervised learning is the preferred method when labeled data is absent or the number of labels is unknown~\cite{Ashfahani2022}. It is ideal for scenarios where learning from the inherent structure of the data is crucial.
Reinforcement learning shines when the model must autonomously determine the most effective means to achieve a goal through rewards~\cite{Li2021a}. This approach empowers the model to learn through interaction and consequences.

In summary, the spectrum of machine learning paradigms -- supervised, semi-supervised, unsupervised, and reinforcement learning -- plays an integral role in both traditional and continual learning, tailoring the choice of approach to the nature of the data and learning objectives.

\subsection{Continual Learning Settings}

Continual learning operates within three fundamental settings, each catering to specific learning demands: task incremental (or instance incremental)~\citep{Feng2020,Mai2021,Zhang2021}, class incremental~\citep{Masana2020,Belouadah2021}, and domain incremental (or attribute incremental)~\citep{Daum1993,Ben-David2010,Mathur2019,Kalimuthu2019}. Task incremental refers to the model's ability to train on an infinite number of new tasks as they appear, such as classifying patterns in data~\citep{Maltoni2019}. It is important to note that task incremental data arrive sequentially, distinct from multitask learning, where multiple tasks are learned in parallel. This distinction applies in continual and non-continual learning contexts~\cite{Zhang2021}. Class incremental learning extends the concept of task incremental learning by addressing situations where an unknown number of classes must be accommodated as the number of tasks increases~\cite{Hao2019}. Domain incremental learning reflects a model's ability to continually adapt to new attribute domains as classification tasks proliferate. This adaptation involves accommodating the evolving data distribution in new tasks~\cite{Venkataramani2019,Srivastava2021}. Domain adaptation, a subset of machine learning, deals with adapting a model from a source domain to a target domain exhibiting distinct characteristics~\cite{Singh2020}. For instance, it might involve the adaptation of chest X-ray radiography classification to head X-ray classification. When domain adaptation is performed continually as classification tasks increase, it can be categorized as 'domain incremental'.

\subsection{Continual Learning Methods}

Mitigating catastrophic forgetting in continual learning is a formidable challenge~\cite{Hasselmo2017}. Existing methods (also known as strategies) for minimizing this issue can be categorized into two primary approaches: regularization-based and memory-based methods.

\textit{Regularization-based} methods aim to reduce catastrophic forgetting by applying weight regularization techniques. In a task incremental setting, elastic weight consolidation (EWC) is used, which regularizes network parameters by penalizing changes based on their relevance to prior tasks~\cite{Aich2021}. However, EWC's efficacy diminishes in class incremental and domain incremental scenarios due to gradient instability, making it more suitable for task incremental contexts.
In contrast, learning without forgetting (LwF) uses data regularization to introduce task-specific parameters, optimizing the performance for new and previously learned tasks when new tasks are introduced in continual learning~\cite{Li2018}. LwF excels on small datasets in class incremental scenarios but fares poorly in domain incremental settings, especially when the data distribution varies significantly between prior and current tasks~\cite{Masana2020}.

\textit{Memory-based} methods rely on maintaining episodic memory, a subset of observed examples from previous tasks. The gradient episodic memory (GEM) technique effectively addresses catastrophic forgetting in task incremental scenarios by limiting the size of the episodic memory, controlling how much can be stored~\cite{Lopez-Paz}. On the other hand, GEM proves less effective in class incremental and domain incremental contexts.
Greedy sampler and dumb learner (GDUMB) adds new samples to a memory buffer, and during inference, it retrains the model from scratch using the stored samples~\cite{Prabhu2020}. GDUMB excels with modest datasets and a large memory buffer, delivering optimal performance. However, it may not be the most efficient choice for inference, as it requires retraining from the entire memory buffer. This approach is less effective in domain incremental settings, primarily due to its memory update strategy~\cite{Mai2021}.

\section{Related work}\label{lbl:relatedwork}

In this section, we discuss relevant literature within the scope of this study, spanning three key areas: traditional deep learning for COVID-19 detection (i.e. non-continual deep learning), continual learning for medical images, and mobile applications for COVID-19 screening.

\subsection{Deep Learning Models for COVID-19 Detection} \label{lbl:related-work:non-continual-deep-learning}

Deep neural networks have demonstrated remarkable success in disease classification using radiography images \citep{shen2017deep}. Chest X-rays serve as a common diagnostic tool for assessing lung diseases, and the application of deep learning has significantly mitigated human errors while expediting the diagnostic process, ultimately reducing misdiagnoses and enhancing result turnaround times \citep{Bhandary2020}. Numerous studies have explored using machine learning and deep learning techniques for COVID-19 detection in radiography \citep{Nour2020,Chowdhury2020,Vaishya2020,subramanian2022review}. The body of research in this domain is substantial, focusing on deep learning techniques for detecting COVID-19, as evidenced by many review articles~\citep{shoeibi2024automated,Panday2020,sun2022performance,dong2020role,alghamdi2021deep,aggarwal2022covid,bhosale2023application,alyasseri2022review,sailunaz2023survey,khan2023survey}. These studies can be categorised into those using well-established deep learning architectures or those proposed custom-designed architectures. 

Deep learning architectures such as ResNet, VGG, MobileNet, DenseNet, EfficientNet, and GoogleNet have been harnessed for the classification of COVID-19 chest X-rays~\citep{Punia2020,Chowdhury2020,Sethy2020,Oh,chowdhury2021ecovnet,Abbas2021,Sethy2020,Wang2020,Basu2020,Panwar2020,Rajaraman2020,Saiz2020,Waheed2020,Vaid2020,Brunese2020,Heidari2020,Apostolopoulos2020,Civit-Masot2020,Li2020b,Togacar2020,Apostolopoulos2020,Apostolopoulos2020b,Karim2020,Kassania2021,Ezzat2021,Shelke2021,Luz2022,Ozkaya2020}. Furthermore, custom deep learning architectures have been proposed for classifying COVID-19 images, including architectures like PDCOVIDNet \citep{Chowdhury2020a}, CovMUNET \citep{Sayyed2020}, COVID-SDNet \citep{Tabik2020}, COVID-Net \citep{Wang2020a}, CovXNet \citep{Mahmud2020}, CoroNet \citep{Khan2020}, COVID-CAPS \citep{Afshar2020}, DarkCovidNet \citep{Ozturk2020} and Convolutional capsnet \citep{Toraman2020}. While these custom architectures have demonstrated high accuracy in COVID-19 chest X-ray classification, our approach focuses on using well-established models for their proven reliability and effectiveness as foundational models in continual learning scenarios.
Moreover, these well-established models offer several advantages for continual learning. Firstly, they come with pre-trained weights from large and diverse datasets, which can be fine-tuned for specific tasks like COVID-19 detection, thereby accelerating the training process. Secondly, their proven architectures ensure a stable and efficient baseline, which is crucial when implementing continual learning strategies that require incremental updates without compromising the integrity of previously learned information.

\subsection{Continual Learning for Medical Images}
Continual learning has seen increasing application in the medical domain~\citep{Lee2020,Venkataramani2019}, particularly for tasks such as the classification of chest X-rays and CT scans~\citep{pmlr-v121-lenga20a}.
Continual learning classification solutions such as elastic weight consolidation (EWC) and learning without forgetting (LwF) have been deployed alongside a foundational DenseNet architecture. These combinations have been used for training and classifying chest X-rays from patients with various conditions, including normal cases, pneumonia, and other diseases, excluding COVID-19~\citep{pmlr-v121-lenga20a}. In this context, EWC and LwF were compared against joint training, with joint training demonstrating superior results in the final evaluations. However, it is important to note that joint training does not adhere to the principles of continual learning and deviates from the established continual learning setup, as it considers all data in the task sequence simultaneously \citep{Delange2021}. Consequently, joint training will not be regarded as a continual learning method for this research.

Additionally, continual learning has been used in domain incremental machine learning for X-ray segmentation~\citep{Srivastava2021} and segmentation tasks in medical radiography more broadly~\citep{Gonzalez}. These approaches have shown promise in handling evolving data distributions and adapting to new tasks within the medical imaging domain.

However, to our knowledge, there has been no study for a continual learning model specifically designed to classify COVID-19. This paper seeks to address this gap by exploring the application of continual learning to COVID-19 classification, leveraging the strengths of well-established models and continual learning strategies to provide a robust and adaptive solution for this critical medical challenge.

\subsection{Mobile Apps for COVID-19 Screening}

In response to the COVID-19 emergency, mobile applications have been increasingly used for healthcare delivery and COVID-19 screening \citep{Alanzi2021}. Systematic literature reviews have demonstrated a substantial body of work on COVID-19 mobile applications \citep{Kondylakis2020}. Some studies have explored applications designed to assist in detecting COVID-19 symptoms using device sensors. For instance, the COVID-19 CheckUp application uses an algorithm that utilizes temperature sensors, heart rate monitors, and the patient's body's reaction to medication to detect symptoms of pneumonia or COVID-19 \citep{Heo2020}. These applications also have features that could benefit from implementing artificial intelligence. In some cases, researchers have investigated the potential of screening for COVID-19 by analyzing patients' speech and coughing sounds \citep{Brown20,Imran2020}. They collected vocal data from patients and trained models to recognise COVID-19 based on sound patterns. Mobile phone-based web surveys have also been explored for early diagnostic purposes \citep{SrinivasaRao2020}. Moreover, mobile artificial intelligence applications have been developed for the classification of COVID-19 in chest X-ray images~\citep{Li2020b}. These applications leverage the power of AI to enhance diagnostic accuracy and speed. 

However, to our knowledge, no application is known to use continual machine learning for COVID-19 screening from chest X-rays, as presented in this study. This paper aims to fill this gap by proposing and evaluating a novel continual learning-based mobile app for COVID-19 screening, providing an adaptive and efficient solution for ongoing pandemic challenges.

\section{Methodology}\label{lbl:methodology}
\begin{figure}[htbp]
    \centering
    \includegraphics[width=1\textwidth]{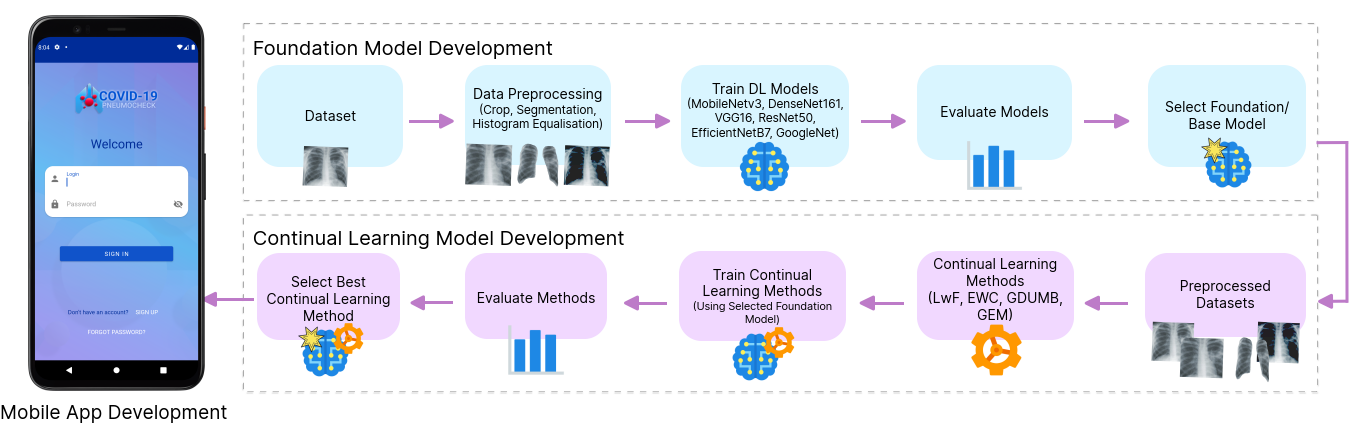}
    \caption{An overview of our methodology.}
\label{fig:methodology_for_continual_learning_model}
\end{figure}

Figure~\ref{fig:methodology_for_continual_learning_model} provides an overview of our methodology for developing the continual learning model for COVID-19 detection. The process begins with data preprocessing, which involves applying various image preprocessing strategies (such as cropping, segmentation and
histogram equalization) to assess their effects on the performance of a deep learning model. This step is followed by training and evaluating state-of-the-art deep learning architectures to identify the most suitable foundation model. The selected foundation model is then subjected to various continual learning methods. These methods are trialed and rigorously evaluated to determine the best performing continual learning model for COVID-19 detection.

\subsection{Foundation Model} \label{sec:foundation_model}

In this study, we first needed to create a foundation model that would serve as the base for our continual learning model. To achieve this, we analyzed existing research and identified several architectures as potential candidates for a robust foundation model. The candidate architectures were: MobileNetV3, ResNet50, VGG16, DenseNet161, EfficientNetB7, and GoogleNet. These architectures were chosen based on their proven effectiveness in COVID-19 classification, as discussed in Section \ref{lbl:related-work:non-continual-deep-learning}.

We trained and evaluated a model from each of these architectures on a dataset of COVID-19 chest X-rays to determine the best performing foundation model. The objective was to identify the most suitable foundation model by comparing the performance of different architectures.
In addition to evaluating different architectures, we also experimented with three different image preprocessing techniques: image cropping, segmentation, and equalization~\citep{caseneuve2021chest}. These techniques were tested to see if they had any impact on the performance of the models. Image cropping involves focusing on the region of interest, segmentation aimed at isolating relevant features, and equalization improves the contrast of the images~\citep{rahman2021exploring}.

Our approach involved applying these state-of-the-art architectures in a supervised classification setup, which is a conventional method in deep learning for sequential tasks. By doing so, we aimed to establish a solid foundation model that could then be used effectively in our continual learning framework.
The experiments and evaluations helped us identify the best model that balances high accuracy with computational efficiency, providing a strong starting point for developing our continual learning model.

\subsection{Continual Learning Methodology}

To develop and evaluate the continual learning model, we first needed to identify and choose the best-performing COVID-19 continual learning model. A task incremental continual model was built from each foundational model architecture to find a candidate continual learning model for the mobile application. Task incremental was chosen over class incremental because the COVID-19 continual learning model required classes to increase during training or as a reference for the model to be effective. Since the data only had three classes to consider (COVID-19, pneumonia, and normal), task incremental learning was more appropriate.

The continual learning model was developed using a supervised learning paradigm in a task incremental setting. We created a continual learning model using each foundational model architecture: MobileNetV3, ResNet50, VGG16, DenseNet161, EfficientNetB7, and GoogleNet. These models were trained with four continual learning methods. We considered Elastic Weight Consolidation (EWC) and Learning without Forgetting (LwF), as they are popular methods that have shown good results in mitigating forgetting in related works. Additionally, we considered GDUMB and Gradient Episodic Memory (GEM) from memory-based solutions.

The training used a supervised paradigm, which is well-suited for the classification of medical imagery with a given set of labeled images. In a supervised paradigm, experts (in this case, doctors) label the X-ray images as they appear and train the model with diagnostics from an official test (such as an RT-PCR test) as the source of truth~\citep{Dou2017}. Patients may also submit RT-PCR test results, avoiding the need for a doctor's diagnosis.

Supervised learning is ideal for COVID-19 images because the classes are known (e.g., COVID-19, Pneumonia, and Normal) and the labeled data is provided by experts. The main advantage of supervised learning is that the data are not weakly labeled but have been carefully labeled by medical experts. This ensures high-quality training data, leading to better model performance

We aimed to identify the best performing continual learning model for COVID-19 detection by evaluating these combinations of foundational architectures and continual learning methods. This model would then be integrated into the mobile application to provide a robust and adaptive tool for ongoing COVID-19 screening and diagnosis.

\subsection{COVID-19 Screening Mobile App}
A COVID-19 screening mobile app needs to accommodate patients, doctors, and researchers to be a useful tool. To implement and expand on existing related works, the app's core functionality must include the ability to classify X-rays, provide a prediction that can be used as a prediagnosis, and gather COVID-19 symptoms from patients.

\textit{Design Considerations and Requirements:}

\begin{itemize}
    \item \textit{Data Input and Continual Learning}: The application must facilitate data input into the system and cope with new data. Continual learning is essential to adapt the model to real-life clinical changes based on information gathered from users.
    \item \textit{User Management}: The app must include a module for authenticating and managing users. Patients, doctors, and researchers should be able to log in, submit chest X-rays or symptoms, and access relevant features.
    \item \textit{AI-Powered pre-diagnosis}: Artificial intelligence should be leveraged to assist in pre-diagnosing user submissions. This includes providing predictions based on an X-ray image and reported symptoms.
    \item \textit{Local and Network-Based Models}: The app requires a local deep learning model for on-device recognition when a network connection is unavailable. This ensures the app remains functional even without internet access.
    \item \textit{Adaptability to New Strains and Symptoms}: As new strains of COVID-19 arise, the application should adapt and learn from changes in symptoms reported by users, maintaining accurate predictions of infection. The app should continuously update its model based on new data.
    \item \textit{Symptom Gathering}: The app should gather patient symptoms through surveys, sensors, and input data, recording and screening the illness trajectory.
\end{itemize}

\textit{Functionality for Different Users}:

\begin{itemize}
    \item \textit{Patients}: Patients can submit chest X-rays, report symptoms, and receive prediagnosis predictions. They can also update their health status and track symptom changes over time.
    \item \textit{Doctors}: Doctors can review patient submissions, provide diagnostic feedback, and update patient records. They can also use the app to monitor patient progress and adjust treatment plans.
    \item \textit{Researchers}: Researchers can access anonymized data for analysis, track trends in symptom reports, and contribute to model training by verifying predictions and outcomes.
\end{itemize}

\textit{Goals and Evaluation:}

\begin{itemize}
    \item \textit{Design and Development}: The goal is to design, develop, and implement an application that meets these requirements. The app should provide a seamless user experience, ensuring ease of use and reliability.
    \item \textit{Performance Evaluation}: The application's performance will be evaluated while running the continual learning model and the locally deployed foundation model. This evaluation will include accuracy, adaptability to new data, and the effectiveness of the prediagnosis predictions.
\end{itemize}

By incorporating these features and design considerations, the COVID-19 screening mobile app aims to provide a comprehensive tool for early detection and monitoring, contributing to better healthcare outcomes during the pandemic.

\section{Experimental evaluation}\label{lbl:experiment}

The objectives of the experimental evaluations are: (1) assessing the effect of various dataset preprocessing strategies on the performance of the foundation model, (2) evaluating the state-of-the-art deep learning architectures to identify the best performing  foundation model to be used in continual learning, and (3) evaluating the performance of different continual learning methods in detecting COVID-19. A CUDA enabled RTX 3080 Ti Desktop GPU with 12GB of memory was used for training and evaluation.

\subsection{Dataset}
\begin{figure}[!ht]
    \centering
    \includegraphics[width=.65\textwidth]{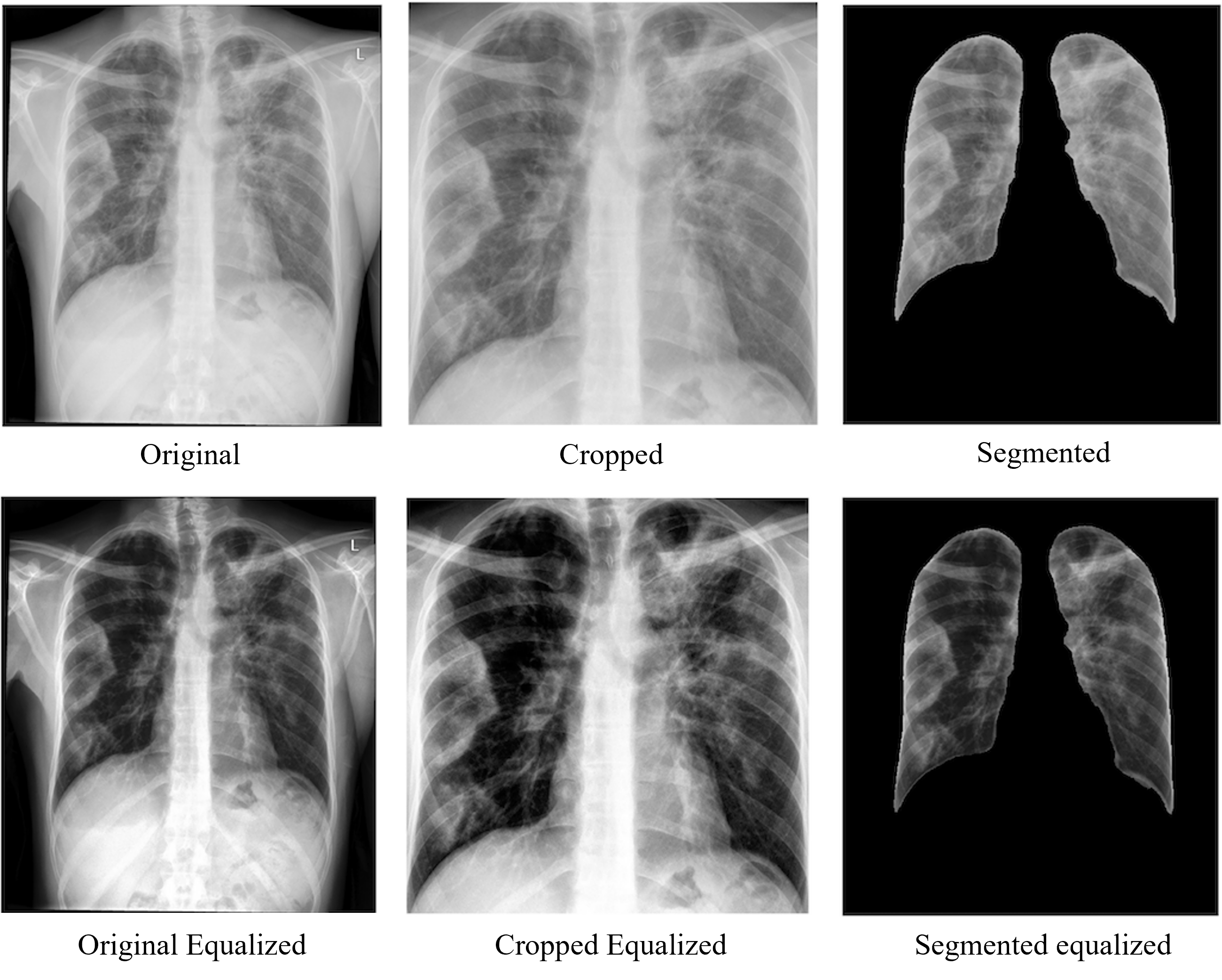}
    \caption{In addition to the original unaltered dataset, cropping and lung segmentation were used as preprocessing strategies. Another set of images was generated by individually applying histogram equalization to each preprocessing strategy.}
    \label{fig:preprocessing_strategies_sample}
\end{figure}

For experimental evaluation, we created our dataset by merging two COVID-19 X-ray image repositories: COVID-19 Posterior-Anterior Chest Radiography Images (X-rays)~\citep{Sait2021} and COVID-19 Radiography Database~\citep{Rahman2021}. All images were resized to $224\times224$ to accommodate the input size of the deep learning foundation models. We applied various image preprocessing strategies (such as cropping, segmentation and histogram equalization) to assess their effects on the performance of a deep learning model. Figure \ref{fig:preprocessing_strategies_sample} shows samples of different preprocessing techniques on COVID-19 chest X-ray images. 

Table~\ref{tab:dataset_preprocessing_strategies} presents six versions of COVID-19 datasets created by applying various preprocessing strategies. Cropping was done by center-cropping the image to focus on the lung area~\citep{Li2020a}, removing irrelevant parts such as arms, neck, shoulders, black surrounding areas and stomach areas, and then resizing the image to the model requirements. The dataset was segmented using a pre-trained ResNet34 model~\citep{RenatAlimbekovIvanVassilenko} with dice evaluation results of 0.9657. We used the X-ray segmentation library~\citep{RenatAlimbekovIvanVassilenko} to segment the left and right lungs by masking and cropping. Any other area without a lung was set to a pixel value of 0. The resulting image was the left and right lungs on a black background. The same image test data set has been used consistently across the foundation model testing and continual learning testing. 

\begin{table}[!ht]
    \centering
    \caption{Dataset Preprocessing Strategies}
    \label{tab:dataset_preprocessing_strategies}
    \begin{tabular}{clcccc}
        \hline
        No. & Preprocessing strategies & Classes & Train & Validation & Test \\
        \hline
        \hline
        1-1 & Orginal & \multirow{6}{*}{COVID-19, Normal, Pneumonia} & \multirow{6}{*}{10800} & \multirow{6}{*}{1200} & \multirow{6}{*}{1500}\\
        1-2 & Original equalized &&&&\\
        2-1 & Cropped &&&&\\
        2-2 & Cropped equalized &&&&\\
        3-1 & Segmented &&&&\\
        3-2 & Segmented equalized &&&&\\
        \hline
        
        \hline
    \end{tabular}
\end{table}

\subsection{Results for Foundation Model}
For the foundation model experiment, we used the Adam optimizer and ran training for 70 epochs with early stopping and saved the model with the highest accuracy. Training batch size was 64 and the learning rate was set to 0.001. The input image size was set to $224\times224$, and pre-trained ImageNet weights were used for all the architectures. 

All models were trained and tested with Dataset No. 1-1 (the original data set without preprocessing). Table \ref{tab:result-1} reports the results. In all metrics, DenseNet161 performed better than the other architectures because it had the best performance in all evaluation metrics. DenseNet161 has a favorable size compared to VGG16 and EfficientNetB7, but MobileNetV3 had the smallest size at 19.7 MB and almost achieved the performance of EfficientNetB7, which is approximately 13 times larger than MobileNetV3. Since accuracy is a priority, we tested the highest performing architecture, DenseNet161, on all dataset preprocessing strategies (equalization, segmentation, and cropping) to identify which technique has the best impact on the most accurate model performance. 

\begin{table}[!htb]
    \centering
    \caption{Performance of different deep learning models for Dataset 1-1 (original)}
    \label{tab:result-1}
    \begin{tabular}{lcccccccc}
    \hline
        DL model & Precision & Specificity & Recall & F1-score & \makecell{Accuracy\\($\pm$95\%CI)} & AUC & \makecell{Model size\\(MB)}  \\
        \hline
        \hline
        MobileNetV3 & 96.33 & 98.17 & 96.33 & 96.33 & 96.33$\pm$0.95 & 0.9962 & 19.7\\
        VGG16 & 92.80 & 96.40 & 92.80 & 92.80 & 92.80$\pm$1.31 & 0.9902 & 545.5\\
        ResNet50 & 93.47 & 96.73 & 93.47 & 93.47 & 93.47$\pm$1.25 & 0.9924 & 98.6\\
        DenseNet161 & \textbf{96.87} & \textbf{98.43} & \textbf{96.87} & \textbf{96.87} & \textbf{96.87$\pm$0.88} & \textbf{0.9979} & 111.7\\
        EfficientNetB7 & 96.47 & 98.23 & 96.47 & 96.47 & 96.47$\pm$0.94 & 0.9968 & 262.2\\
        GoogleNet & 95.87 & 97.93 & 95.87 & 95.87 & 95.87$\pm$1.01 & 0.9953 & 24.7\\
        \hline
        
        \hline
    \end{tabular}
\end{table}

Table \ref{tab:evaluation-results-densenet161-all} shows the results for the DenseNet161 architecture on all dataset preprocessing strategies. DenseNet161 performed best with the original dataset without equalization. The segmented dataset showed the worst results but improved when the images were equalized. Segmentation was the only strategy that showed improved results when equalization was performed. Equalizing the cropped dataset showed worse results than just performing cropping and equalizing the original dataset showed worse results than the original dataset without any preprocessing. The original unmodified data set showed the highest accuracy and COVID-19 precision of the DenseNet161 model and was the strongest candidate for the preprocessing strategy for the continual learning model.
\begin{table}[!htb]
    \centering
    \caption{Evaluation results of all the preprocessing strategies of the DenseNet161 model}
    \label{tab:evaluation-results-densenet161-all}
    \resizebox{1\textwidth}{!}{
    \begin{tabular}{clccccccccc}
       \hline
       \multirow{2}{*}{Dataset} & \multirow{2}{*}{Class} & \multirow{2}{*}{Precision} & \multirow{2}{*}{Recall} & \multirow{2}{*}{F1-score} & \multicolumn{6}{c}{Overall}\\
       \cline{6-11}
       &&&&&Precision & Specificity & \makecell[c]{Sensitivity/\\Recall} & F1-score & \makecell{Accuracy\\($\pm$95\%CI)} & AUC\\
       \hline
       \hline
       \multirow{3}{*}{1-1} & COVID-19 & 99.38 & 95.80 & 97.56 & \multirow{3}{*}{96.87} & \multirow{3}{*}{96.43} & \multirow{3}{*}{\textbf{96.87}} & \multirow{3}{*}{\textbf{96.87}} & \multirow{3}{*}{\textbf{96.87$\pm$0.88}} & \multirow{3}{*}{\textbf{0.9979}}\\
       & Normal & 93.02 & 98.60 & 95.73 &&&&&&\\
       & Pneumonia & 98.57 & 96.20 & 97.37 &&&&&&\\
       \hline
       
       \multirow{3}{*}{1-2} & COVID-19 & 98.56 & 95.6 & 97.06 & \multirow{3}{*}{96.07} & \multirow{3}{*}{98.03} & 
       \multirow{3}{*}{96.07} & 
       \multirow{3}{*}{96.07} & \multirow{3}{*}{96.07$\pm$0.99} & \multirow{3}{*}{0.9966}\\
       & Normal & 91.76 & 98 & 94.78 &&&&&&\\
       & Pneumonia & 98.34 & 94.6 & 96.43 &&&&&&\\
       \hline
       
       \multirow{3}{*}{2-1} & COVID-19 & 97.31 & 94.2 & 95.73 & \multirow{3}{*}{94.27} & \multirow{3}{*}{97.13} & \multirow{3}{*}{94.27} & \multirow{3}{*}{94.27} & \multirow{3}{*}{94.27$\pm$1.18} & \multirow{3}{*}{0.9931}\\
       & Normal & 90.29 & 94.8 & 92.49 &&&&&&\\
       & Pneumonia & 95.52 & 93.8 & 94.65 &&&&&&\\
       \hline
       
       \multirow{3}{*}{2-2} & COVID-19 & 97.67 & 92.4 & 94.96 & \multirow{3}{*}{92.87} & \multirow{3}{*}{96.43} & \multirow{3}{*}{92.87} & \multirow{3}{*}{92.87} & \multirow{3}{*}{92.87$\pm$1.31} & \multirow{3}{*}{0.9897}\\
       & Normal & 88.21 & 92.8 & 90.45 &&&&&&\\
       & Pneumonia & 93.21 & 93.4 & 93.31 &&&&&&\\
       \hline
       
       \multirow{3}{*}{3-1} & COVID-19 & 97.13 & 88 & 92.34 & \multirow{3}{*}{92.13} & \multirow{3}{*}{96.07} & \multirow{3}{*}{92.13} & \multirow{3}{*}{92.13} & \multirow{3}{*}{92.13$\pm$1.36} & \multirow{3}{*}{0.9897}\\
       & Normal & 88.78 & 91.8 & 90.27 &&&&&&\\
       & Pneumonia & 91.13 & 96.6 & 93.79 &&&&&&\\
       \hline
       
       \multirow{3}{*}{3-2} & COVID-19 & 98.24 & 89.2 & 93.5 & \multirow{3}{*}{92.87} & \multirow{3}{*}{96.43} & \multirow{3}{*}{92.87} & \multirow{3}{*}{92.87} & \multirow{3}{*}{92.87$\pm$1.31} & \multirow{3}{*}{0.9896}\\
       & Normal & 89.49 & 92 & 90.73 &&&&&&\\
       & Pneumonia & 91.54 & 97.4 & 94.38 &&&&&&\\
       \hline
       
       \hline
    \end{tabular}
    }
\end{table}

\subsection{Results for Continual Learning}

In this section, we compare the various continual learning methods when applied to the DenseNet161 model, as this foundation model architecture had the best overall performance. After each training experience, we provide the average results of the evaluation, including average accuracy, average forgetting, overall performance measure, and average inference time.

Accuracy was recorded for every experience. An experience is considered a scenario in which the model has to train on a subset of unseen images as they come, which simulates a real-world experience or scenario in which information flows in continually from the real world. The average accuracy measures the accuracy of the model throughout its evaluated life of experiences. In this experiment, there were 25 experiences.

Forgetting was calculated by capturing the highest level of catastrophic forgetting within the model's life of 25 experiences. To calculate the maximum catastrophic forgetting, we measure the highest forgetting per experience and obtain the average maximum forgetting of all experiences~\citep{Chaudhry2018}. The forgetting performance is considered most favorable when forgetting is as low as possible, ideally closest to -1.

To consider both accuracy and forgetting, an evaluation metric for overall performance was created. Overall performance is obtained by the formula below:
\begin{equation}\label{equ:overallperformance}
   p = \frac{1}{2} (\bar{a} + \frac{100 -  \bar{f}}{2})
\end{equation}
Where \(p\) is overall performance calculated by the equal weighted average accuracy plus normalized average forgetting. The average accuracy is defined as \(\bar{a} \in [0, 100]\) and average forgetting is defined as \(\bar{f} \in [-100, 100]\). Here, 100 average accuracy and -100 average forgetting will give an overall performance of 100, which is the best possible performance. On the other hand, 0 accuracy and 100 forgetting will equal 0 overall performance, which is the worst possible performance. We also calculated the average time of performing evaluation after every experience. The evaluation was 1500 images at the end of every experience, averaged across the 25 experiences. 
 
\begin{table}[!htbp]
    \centering
    \caption{Continual Learning Model evaluation results using DenseNet161 regularization and memory-based methods}
    \label{tab:continual-learning-results}
    \begin{tabular}{llcccc}
       \hline
       \multicolumn{2}{c}{Method}  & \makecell[t c]{Avg. accuracy\\($\pm$ Std. Dev.)} & \makecell[t c]{Avg. forgetting\\($\pm$ Std. Dev.)} & \makecell[t c]{Overall\\ performance} & \makecell[t c]{Avg. time\\(ms)}\\
       \hline
       \hline
        \multirow{2}{*}{Regularization based} & LwF & \textbf{94.44$\pm$0.74}
 & \textbf{0.91$\pm$0.91} & \textbf{71.99} & 34.84 \\
        & EWC & 94.28$\pm$1.96 & 1.19$\pm$2.14 & 71.84 & 35.62 \\
        \hline
         \multirow{8}{*}{Memory-based} & GDUMB (k=200) & 93.89$\pm$1.35 & 1.56$\pm$1.48 & 71.56 & 32.84 \\
        & GDUMB (k=1280) & 94.33$\pm$1.09 & 1.13$\pm$1.23 & 71.88 & 35.53 \\
        & GDUMB (k=2560) & 94.18$\pm$1.13 & 1.25$\pm$1.26 & 71.78 & 35.48 \\
        & GDUMB (k=5120) & 94.18$\pm$1.13 & 1.25$\pm$1.26 & 71.78 & 35.08 \\
        & GEM (k=200) & 92.43$\pm$3.62 & 3.82$\pm$3.91 & 70.26 & 32.72 \\
        & GEM (k=1280) & 92.43$\pm$3.62 & 3.82$\pm$3.91 & 70.26 & 35.52 \\
        & GEM (k=2560) & 92.43$\pm$3.62 & 3.82$\pm$3.91 & 70.26 & 35.53 \\
        \hline
        
        \hline
    \end{tabular}

\end{table}
 The results reported in Table \ref{tab:continual-learning-results} show that DenseNet161 performed best with Learning without Forgetting (LwF) achieving 94.44\% accuracy and the lowest forgetting of 0.91. Of the memory-based approaches, the GDUMB method has performed overall better than GEM, as indicated by the higher overall performance of GDUMB. When GDUMB was set at $k = 1280$, it was not only the optimal $k$ size, but with a mean time of 35.53 seconds and an overall performance of 71.88, it performed better and faster than the EWC method based on regularization. LwF had better overall performance than GDUMB ($k=1280$), but interestingly, the average time was longer for LwF. 
 
\section{Mobile app implementation}\label{lbl:app}
\subsection{Mobile App Architecture}

\begin{figure}[!htb]
    \centering
    \includegraphics[width=.8\textwidth]{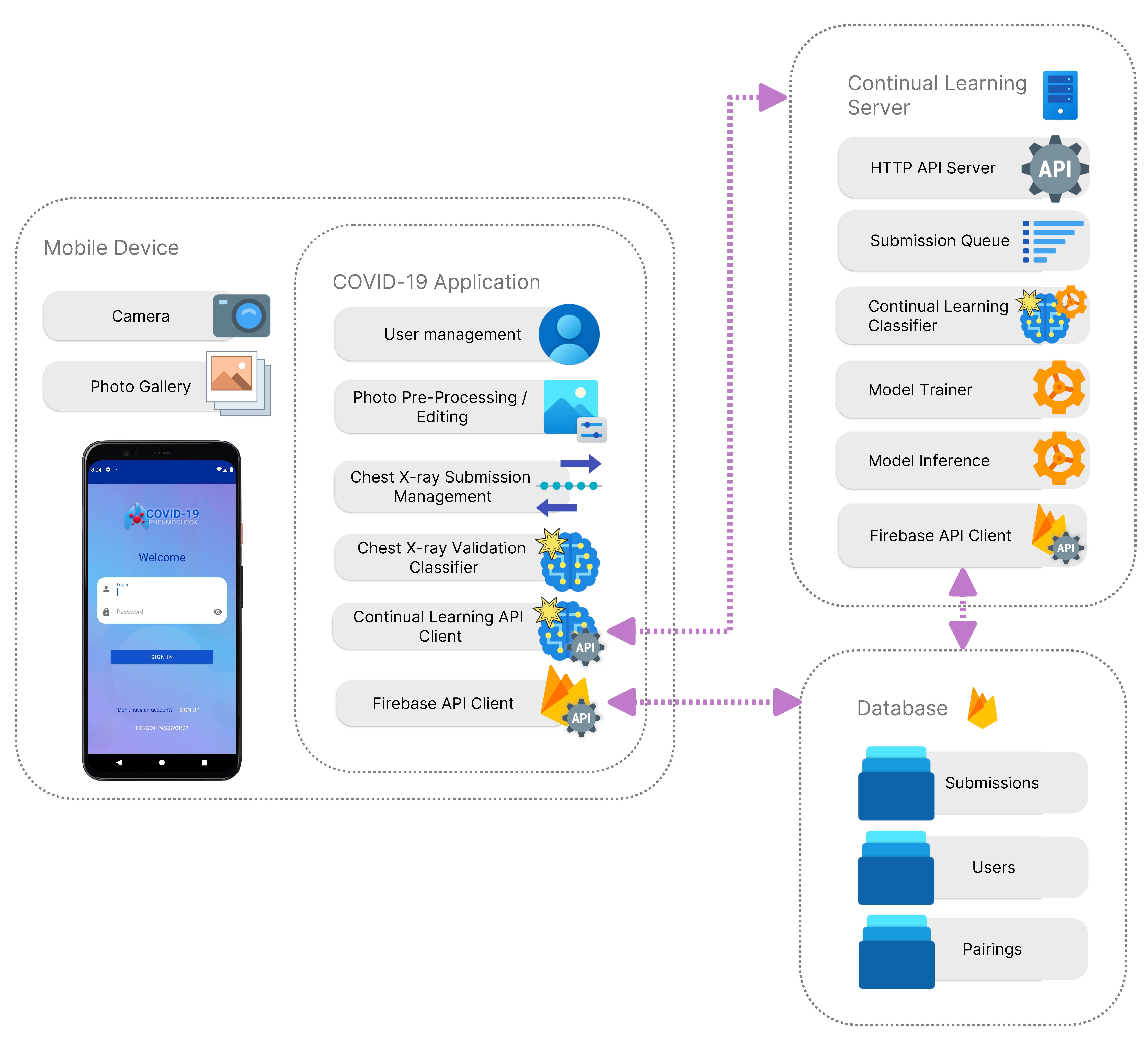}
    \caption{This architecture encompasses a mobile application, continual learning server, and a database.}
    \label{fig:app_continual_learning_system_architecture_details}
\end{figure}

This section presents our mobile app architecture for COVID-19 screening. The architecture comprises three components as depicted in Figure \ref{fig:app_continual_learning_system_architecture_details}, the user-facing mobile application, the database, and the continual learning server.

\paragraph{Mobile Application}The mobile application integrates several modules to enable diverse functions:

\begin{enumerate}
    \item \textit{User Management Module}: This module handles user authentication, login, registration, and profile management, ensuring secure access to the application.
    \item \textit{Photo Preprocessing and Editing Module}: Executes necessary image transformations and enhancements to refine chest X-ray images before analysis.
    \item \textit{Chest X-ray Submission Management Module}: Organizes and manages submitted chest X-ray images, streamlining the screening process.
    \item \textit{Chest X-ray Validation Classifier Module}: Provides automated input validation through machine learning binary classification, distinguishing between valid and invalid chest X-ray images to ensure quality and accuracy.
    \item \textit{Continual Learning Server API Client Module}: Establishes communication with the external Continual Learning server, leveraging the dynamic nature of continual learning to adapt to evolving COVID-19 chest X-ray patterns.
    \item \textit{Firebase API Client Module}: Integrates with Firebase services, providing real-time database capabilities, cloud storage functionalities, and user authentication mechanisms, enhancing the reliability and accessibility of application data.
\end{enumerate}
These modules collectively facilitate user interaction, guarantee data security and privacy, optimize chest X-ray image analysis, and enable evolution in response to the dynamic nature of COVID-19 screening needs. The robust architecture and extensive functionality make the mobile application an effective tool to aid in COVID-19 diagnosis and prevention efforts.

\paragraph{Database}The database securely and efficiently stores user profiles, doctor-patient pairings, and chest X-ray submissions:

\begin{enumerate}
    \item \textit{User Profiles}: Stored in a structured manner with unique identifiers for efficient retrieval and updates.
    \item \textit{Doctor-Patient Pairings}: Maintains records of pairings, ensuring correct associations between patients' X-ray submissions and responsible doctors.
    \item \textit{Chest X-ray Submissions}: Keeps a comprehensive record of all submissions, each associated with unique identifiers and linked to corresponding patients and doctors, ensuring a traceable history of interactions.
\end{enumerate}
This structured and relational storage approach enhances the overall reliability and accessibility of data, supporting the dynamic and evolving needs of the mobile application.

\paragraph{Continual learning server (back-end server)} It includes several key components:

\begin{enumerate}
    \item \textit{Queue Processing Module}: Manages classification and training requests from the mobile app. The module processes one request at a time, determining whether the model should be prepared for training or inference.
    \item \textit{Continual Learning Model}: Examines the submission type when processing a submission image. If the submission type is \verb|classify|, the model performs inference on the image and provides a classification result of either \verb|COVID-19|, \verb|Pneumonia|, or \verb|Normal|, which is then written back to the cloud-hosted database. If the submission type is \verb|learn|, the model expects a label, trains on the submitted chest X-ray image using the provided label, and marks the corresponding entry in the database as \verb|learned|, recording a timestamp for the learning process.
    \item \textit{Database Interaction}: The server interacts with the database through an API client to access submission images and update the database with results from classification or learning processes.
\end{enumerate}
The mobile app's Chest X-ray Submission Management module retrieves the classification results from the cloud-hosted database and displays them to the user. The X-ray submission manager fetches updates from the database and informs the user that the submission confirmation has been learned by the model.

\subsection{Mobile App Design}

\begin{figure}[htbp]
    \centering
    \includegraphics[width=.4\textwidth]{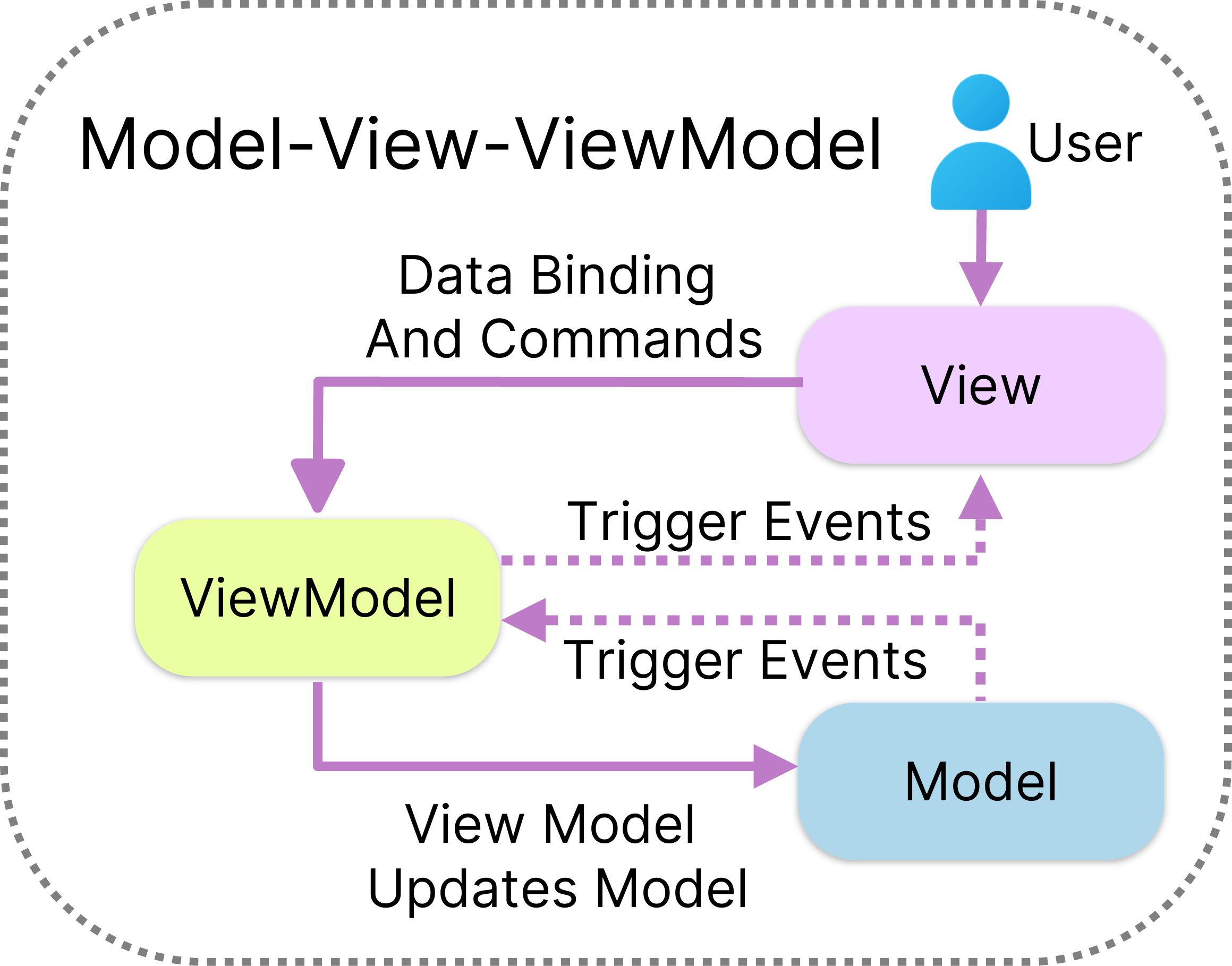}
    \caption{Diagram illustrating the interaction between the View, ViewModel, and Model in the MVVM design pattern.}
    \label{fig:app_design_view_model_viewmodel}
\end{figure}

The architecture of the mobile application's user interface is predicated on the Model-View-ViewModel (MVVM) design pattern, as depicted in Figure \ref{fig:app_design_view_model_viewmodel}. This pattern provides a systematic methodology for decoupling the responsibilities of data representation and business logic, thereby promoting a clean architectural structure and well-organized code. The front-end application implements the MVVM design pattern, specifically integrating the Firebase API client within the ViewModel layer. This integration allows the application to use Firebase's powerful features, such as snapshot listeners and seamless data synchronization. The ViewModel, acting as an intermediary component between the View and the Model, plays a crucial role in this design pattern. It encapsulates the presentation logic and provides the data and operations that the View needs. In our mobile app, the ViewModel interacts with the Firebase API client to communicate with the underlying data model, which is stored in Firebase’s cloud-based database.

Integrating the Firebase API client within the ViewModel layer brings a notable advantage, such as using snapshot listeners. These listeners facilitate real-time updates from Firebase, keeping the ViewModel in sync with any changes in the data stored in the Firebase database. Consequently, the ViewModel responds promptly to changes, updates its internal state, and transmits the updated data to the View. This ensures the user interface remains current, offering a responsive and dynamic user experience.
Moreover, the ViewModel layer efficiently manages the process of inflating views with data. When data is fetched from Firebase, the ViewModel formats and prepares it for easy consumption by the View. This facilitates seamless binding between the data and the UI elements, guaranteeing that the user interface accurately mirrors the latest information from the Firebase database.

In summary, integrating the MVVM design pattern, with the Firebase API client within the ViewModel layer, improves the mobile app's user interface. This strategy encourages separation of concerns, simplifies code organization, enables real-time updates through Firebase's snapshot listeners, and facilitates the efficient population of views with data. By applying these design principles, the mobile app attains a robust and responsive user interface while exploiting the powerful capabilities of Firebase.
\subsection{App Features}

The app aims to serve as an efficient and reliable tool for diagnosing COVID-19 and other respiratory conditions using chest X-ray images. It integrates various features to enhance the user experience and ensure accurate classification of radiography submissions. In this section, we explore the key features of our application, including the X-ray validator, inference logic, training capabilities, and user interface design.

\textit{Chest X-ray Validator Model.} 
To maintain the integrity of the training data and prevent non-X-ray images from being processed, the application features an X-ray validator with two classifiers. \textit{Classifier 1} validates the submitted images to ensure they are valid chest X-ray images, preventing misclassification caused by inappropriate inputs. Once Classifier 1 confirms the image is a valid chest X-ray, \textit{Classifier 2} classifies the images into three categories: `COVID-19', `pneumonia', or `normal'. This step provides valuable insights into the presence of COVID-19 or other respiratory conditions in the submitted X-ray images. 

\begin{figure}[htbp]
    \centering
    \includegraphics[width=0.85\textwidth]{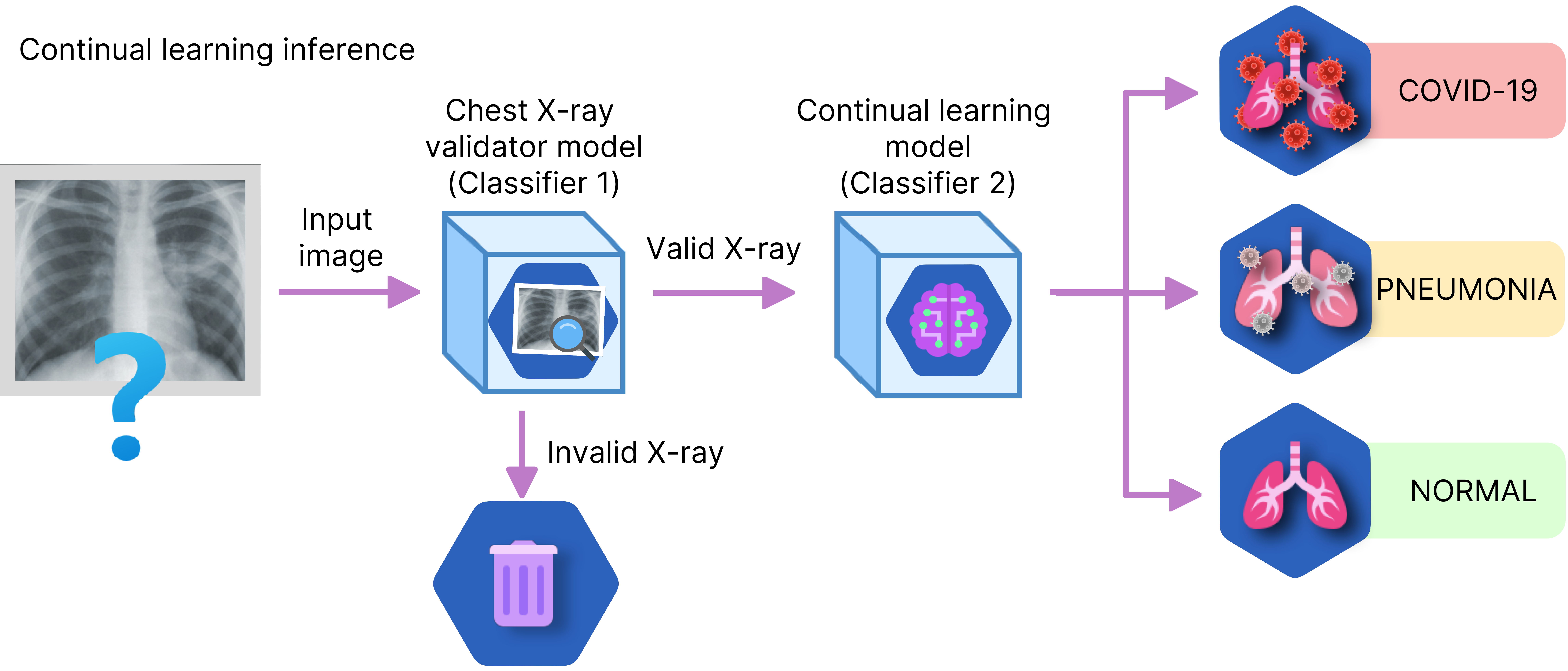}
    \caption{Depiction of the sequential operation of the Chest X-ray validator (Classifier 1) and the inference logic for the continual learning model (Classifier 2), demonstrating the application's approach to image validation and disease classification.}
    \label{fig:classification_inference}
\end{figure}

\textit{Continual Learning Inference Logic.} 
The application’s inference logic, as shown in Figure \ref{fig:classification_inference}, adopts a sequential process to provide timely predictions to the users. When a user submits an X-ray image, the image first undergoes validation by Classifier 1. If the image is deemed valid, it advances to Classifier 2 for further classification. Classifier 2 processes the image and produces a prediction indicating the likelihood of COVID-19, pneumonia, or a normal condition. The pre-diagnosis prediction is immediately communicated to the user, enabling them to take necessary precautions and seek appropriate medical attention.

\begin{figure}[htbp]
    \centering
    \includegraphics[width=0.85\textwidth]{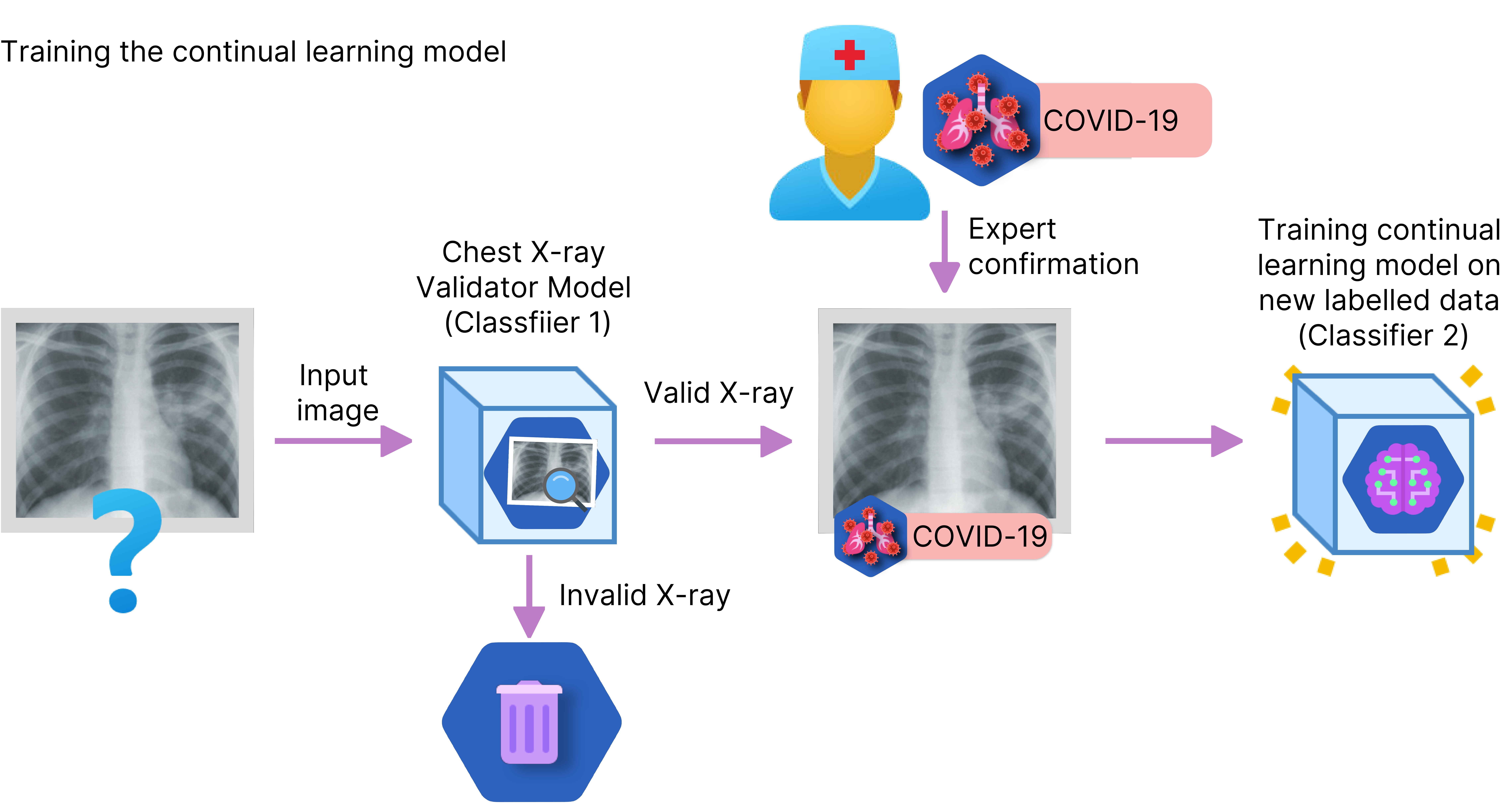}
    \caption{Depiction of the model refinement process through expert confirmation, emphasizing the application's capacity to adapt and improve over time by leveraging real-world clinical data for more accurate future classifications.}
    \label{fig:classification_training}
\end{figure}

\textit{Continual Learning Training Capability.}
To continuously improve classification accuracy, the application uses a continual learning approach. Figure \ref{fig:classification_training} illustrates the training process that involves expert confirmation. When a doctor confirms the prediction provided by the model, the confirmation is submitted to the training module. This expert confirmation acts as ground truth and is used to update the model parameters, boosting its ability to accurately classify future chest X-ray images. Through the incorporation of expert confirmation into the training process, the model benefits from real-world clinical data, enhancing its performance over time and adapting to emerging patterns and variations in chest X-ray images associated with COVID-19 and other respiratory conditions.

\subsection{User Interface Design}

\begin{figure}[htbp]
    \centering
    \includegraphics[width=.82\textwidth]{ui-design-300dpi.pdf}
    \caption{The User Interface of the Mobile Application: Facilitating User-Friendly Interaction and Efficient Image Submission for Rapid Pre-Diagnosis of Respiratory Conditions.}
    \label{fig:app_design_ui}
\end{figure}

The app's user interface has been designed to be intuitive and user-friendly, catering to patients and healthcare professionals. The interface empowers patients to capture images of their chest X-rays using smartphones. Figure \ref{fig:app_design_ui} illustrates the image capture process, facilitating convenient submission of X-ray images by patients. Upon submission of the image, patients receive an immediate pre-diagnosis prediction indicating the likelihood of COVID-19 or other respiratory conditions. This prompt feedback enables patients to take necessary precautions and seek medical advice promptly.

For doctors, the application provides a comprehensive view of the patient’s submission, including the image and the model's prediction. Doctors can use the application to review the predictions and to submit X-ray images for inference, as shown in Figure~\ref{fig:app_design_ui}. The application’s interface allows doctors to confirm or refute the model’s predictions based on their expertise and additional diagnostic tests. This interactive feedback loop enhances the continual learning process, as expert-confirmed cases are used to improve the model’s classification accuracy. 

Collectively, the application’s features such as the X-ray validator, inference logic, training capabilities, and user-friendly interface, enhance its effectiveness as a diagnostic tool for COVID-19 and other respiratory conditions. Through the incorporation of continual learning and expert confirmation, the model adapts and improves over time, boosting its ability to accurately classify chest X-ray images. The application is a valuable resource for patients and healthcare professionals, assisting in timely diagnosis and appropriate management of respiratory diseases.

\subsection{App Evaluation}

The objectives of the app evaluations are: (1) evaluating the efficiency of the continual learning model running on the cloud, and evaluating the app performance based on its (2) efficiency, and (3) resource consumption in different Android devices with various configurations.

\subsubsection{Efficiency of the Continual Learning Model}

The application was set up to use the continual learning model on the cloud and was tested for inference and training times. The model was given 200 chest X-ray images to classify. The time taken for each classification and the overall accuracy were measured. 
The continual learning cloud model’s average inference time was 16.03 milliseconds with a standard deviation of 2.28 milliseconds. The average training time per image was 980.46 milliseconds with a standard deviation of 38.01 milliseconds.

\subsubsection{App Performance}

The local model's inference and loading time were tested in the app running on five different Android devices: Google Pixel 4, Oppo Reno 6, OnePlus 9, Samsung Galaxy S22 and Google Pixel 6. For the inference tests, the model was given 200 different images to classify, and the time taken for the model to perform the forward pass was measured. For the loading tests, the model was loaded 200 times and the loading time for the model to be ready for inference was measured. Table~\ref{tab:model_inference_loading_local} reports the inference time and loading time results. These results indicate that the app's performance across different devices is reasonable. The average loading time ranges from 382.1 ms to 759.9 ms, and the average inference time ranges from 212.4 ms to 665.7 ms, making the app feasible for real-life use.

\begin{table}[htbp]
    \centering
    \caption{Model inference and loading test results}
    \label{tab:model_inference_loading_local}
    \begin{tabular}{@{\extracolsep{4pt}}lcccccc}
    \hline
      \multirow{2}{*}{Phone} & \multicolumn{3}{c}{Loading time (ms)} & \multicolumn{3}{c}{Inference time (ms)} \\
      \cline{2-4}\cline{5-7}
      & Min & Max & Avg ($\pm$ Std. Dev) & Min & Max & Avg ($\pm$ Std. Dev) \\
      \hline
      \hline
       Google Pixel 4 & 379.6 & 383.8 & 382.1$\pm$1 & 367 & 522.2 & 416$\pm$20.8\\
       Oppo Reno 6 & 461.4 & 637.5 & 493.9$\pm$25.3 & 503.3 & 788.5 & 665.7$\pm$49.1 \\
       OnePlus 9 & 352.3 & 360.7 & 355.8$\pm$4.1 & 184.5 & 310.4 & 212.4$\pm$20.1 \\
       Samsung Galaxy S22 & 738.2 & 822.1 & 759.9$\pm$18.9 & 259 & 1289.7 & 468.3$\pm$184.3 \\
       Google Pixel 6 & 418.4 & 451.9 & 430.9$\pm$6.2 & 363.3 & 642.8 & 419.3$\pm$41.2 \\
       \hline
       
       \hline
    \end{tabular}

\end{table}

\subsubsection{App Resource Consumption}

We also tested the resource consumption for the application when running the core functionality of the application. Both the local model inference test and cloud inference and training tests were done. The resource consumption test was done for five minutes and consumption peaks were measured every 10 seconds. The activities performed during performance testing were the login, moving between tabs in the app, classifying images, and confirming classification. CPU, RAM and Energy use were recorded from the Android Studio's App Profiler. Google Pixel 4 was the mobile device used for the resource consumption tests of the mobile application. For the local model inference, the peak CPU average was 20.7\%, the peak RAM average was 369.93 MB, and the battery use was distributed as light, medium and heavy in 70\%, 30\% and 0\% of the time of the test duration, respectively. For the cloud continual learning model inference and training, the maximum CPU average was 14.1\%, the maximum RAM average was 271.56 MB and the battery usage was light, medium and heavy in 96.67\%, 3.33\% and 0\% of the time of the test duration, respectively.

\section{Conclusion and future work}\label{lbl:conclusion}

In this paper, we implemented and evaluated continual learning models, and designed and developed a mobile app, for detecting COVID-19 from chest X-rays. Among the various architectures considered, DenseNet161 emerged as the best model for classifying X-rays using a conventional sequential approach. When DenseNet161 was used as the foundation model for a continual learning approach, the Learning without Forgetting (LwF) method demonstrated the best performance. This finding indicates that a regularization-based solution is the most effective for the chest X-ray continual classification problem based on overall performance. The mobile application also performed better when using the cloud continual learning method, with the added capability to train from labeled data on demand.

There are several avenues for further research within the scope of this study. One potential direction is implementing and expanding the continual learning approach for CT scans and evaluating its performance. Conducting clinical tests using the mobile application in a live environment may refine the application to be considered a digital therapeutic \citep{DigitalTherapeuticsAlliance2020,DigitalTherapeuticsAlliance2021,DigitalTherapeuticsAlliance2019}. Additionally, the mobile application could benefit from incorporating class incremental continual learning to expand its functionality beyond COVID-19. 

\section*{Declaration of competing interest}
The authors declare that they have no known competing financial interests or personal relationships that could have appeared to influence the work reported in this paper.

\section*{CRediT authorship contribution statement}
\textbf{Danny Falero:} Conceptualization, Methodology, Formal analysis, Software, Data Curation, Writing - Original Draft, Visualization.
\textbf{Muhammad Ashad Kabir:} Conceptualization, Methodology, Writing - Review \& Editing, Validation, Supervision, Project administration. \textbf{Nusrat Homaira:} Validation, Writing - Review \& Editing.

\section*{Funding}
This research did not receive any specific grant from funding agencies in the public, commercial, or not-for-profit sectors.

\section*{Data availability statement}
The datasets used in this article are openly available at \url{https://data.mendeley.com/datasets/9xkhgts2s6/4} and \url{https://www.kaggle.com/datasets/tawsifurrahman/covid19-radiography-database}.
\bibliographystyle{elsarticle-num-names}
\bibliography{cas-refs}





\end{document}